\documentclass[reprint, superscriptaddress, secnumarabic, amssymb, nobibnotes, aps, prl]{revtex4-1}

\usepackage{graphicx}
\usepackage{epstopdf}
\usepackage[T1]{fontenc}
\usepackage[latin9]{inputenc}
\usepackage{amsbsy}
\usepackage{gensymb}
\begin{document}

\title{Magnetization reversal, giant exchange bias effect and magnetoresistance in oxygen vacancy ordered Sr$_{4}$Fe$_{3}$CoO$_{11}$}

\author{Prachi Mohanty}
\affiliation {Indian Institute of Science Education and Research Bhopal, Bhopal, 462066, India}
\author{Sourav Marik}
\affiliation {Indian Institute of Science Education and Research Bhopal, Bhopal, 462066, India}
\affiliation {CNRS, Universit\'e de Bordeaux, ICMCB, 87 avenue du Dr. A. Schweitzer, Pessac, F-33608, France}
\author{C. Madhu}
\affiliation {CNRS, Universit\'e de Bordeaux, ICMCB, 87 avenue du Dr. A. Schweitzer, Pessac, F-33608, France}
\author{D. Singh}
\affiliation {Indian Institute of Science Education and Research Bhopal, Bhopal, 462066, India}
\author{O. Toulemonde}
\affiliation{CNRS, Universit\'e de Bordeaux, ICMCB, 87 avenue du Dr. A. Schweitzer, Pessac, F-33608, France}
\author{Ravi P. Singh}
\email{rpsingh@iiserb.ac.in}
\affiliation {Indian Institute of Science Education and Research Bhopal, Bhopal, 462066, India}

\date{\today}
\begin{abstract}
\begin{flushleft}
\end{flushleft}

 We report the structural, magnetic, exchange bias and magnetotransport effect in Sr$_{4}$Fe$_{3}$CoO$_{11}$. The material crystallizes in the orthorhombic $\textit{Cmmm}$ space group. It shows antiferromagnetic (G-type) transition (T$_{N}$ = 255 K) along with interesting temperature induced magnetization reversal (T$_{Comp.}$= 47 K measured at 100 Oe). The magnetic reversal can be elucidated considering the increased magnetocrystalline anisotropy with Co substitution. Magnetoresistance measurements shows an interesting crossover from negative to positive side at $\sim$ 100 K. The negative magnetoresistance reaches 80 $\%$ at 25 K in 7 T magnetic field. Giant exchange bias effect is observed below T$_{N}$ under field cooling condition. The origin of the negative magnetoresistance and giant exchange bias in this sample can be attributed to the magnetic frustration.

\end{abstract}
\maketitle
\section{Introduction}

Strongly correlated transition metal oxide (TMOs) material systems has attracted widespread attention as a plethora of exotic phenomena such as high-temperature superconductivity (SC, cuprates) \cite{1}, colossal magnetoresistance (manganites) \cite{2}, exchange bias \cite{3,4} and multiferroicity (bismuth compounds) \cite{5} arises from correlated electron physics in 3d TMOs. Indeed, the strong correlation among electrons is a connecting feature of TMOs exhibiting extraordinary electronic and magnetic properties, and that can develop into a promising technological application (such as in electronics and spintronic field). At the same time, purely by changing the valence state of the transition metal ions, unusual physical properties can emerge in TMOs  This is realized to be one of the critical issues in understanding the exotic properties observed in high-temperature superconductivity\cite{6}, multiferroics \cite{7}, and magnetoresistive compounds \cite{8}. 

Among transition metals, iron is a very attractive transition metal element due to its potential to exhibit several oxidation states and spin states in different environments which often induce interesting properties. Oxygen deficient strontium ferrites are such systems displaying valence changes. In recent years a great number of studies has been done due to their diverse structural and physicochemical properties relevant for potential applications \cite{9,10,11,12,13,14,15}. They belong to a class of anion-deficient perovskite-like compounds, which are known for the Sr$_{n}$Fe$_{n}$O$_{3n-1}$ (or SrFeO$_{3-\delta}$) system. This system is well recognized to display a rich phase diagram depending on the oxygen content, ranging from 3.0 to 2.0, which includes cubic SrFeO$_{3}$\cite{14}, tetragonal Sr$_{8}$Fe$_{8}$O$_{23}$ (SrFeO$_{2.87}$) \cite{9}, orthorhombic Sr$_{4}$Fe$_{4}$O$_{11}$ (SrFeO$_{2.75}$) \cite{11}, Sr$_{2}$Fe$_{2}$O$_{5}$ (SrFeO$_{2.5}$) \cite{16} and SrFeO$_{2}$ phase \cite{15}. The square planer oxygen deficient end member exhibits Fe$^{2+}$ cations with an above room temperature magnetic ordering \cite{15}. The other end phase cubic SrFeO$_{3}$ possesses an unusually high Fe$^{4+}$ oxidation state \cite{14}. The intermediate member Sr$_{4}$Fe$_{4}$O$_{11}$ ( orthorhombic 2$\sqrt{2}$a$_{p}$ $\times$ 2a$_{p}$ $\times$ $\sqrt{2}$a$_{p}$, a$_{p}$ = lattice parameter of the cubic perovskite sub-cell) have an equal number of tetravalent and trivalent iron. Also, Sr$_{4}$Fe$_{4}$O$_{11}$ shows a variety of interesting properties, such as anomalous thermoelectric power and unusual magnetic behavior \cite{11,13}.  Hitherto, all the structural studies have indicated that Fe ions alternatively occupy square pyramidal and octahedral sites and undergo a charge ordering below 675 K with an equal amount of Fe$^{4+}$ and Fe$^{3+}$ alternatively arranged in the lattice \cite{11,12,13,17,18,19}. It is reported that one Fe site (Fe2, Octahedral) forms a long-range AFM order (T$_{N}$ = 240 K), whereas the other (Fe1, square pyramidal) forms a magnetically frustrated spin-glass-like configuration below T$_{N}$, which originate unusual magnetic properties in Sr$_{4}$Fe$_{4}$O$_{11}$.

In this paper, we have shown the structure, complex magnetic and transport properties of the oxygen-deficient perovskite-related compound with composition Sr$_{4}$Fe$_{3}$CoO$_{11}$. Giant exchange bias effect and a sizable magnetoresistance are observed below the magnetic transition of the material. Detailed structural and complex magnetic properties are presented and discussed here.

\section{Experimental Details}

Polycrystalline samples of composition Sr$_{4}$Fe$_{3}$CoO$_{11}$ were synthesized by standard solid state reaction method using stoichiometric amount of high purity SrCO$_{3}$, Fe$_{2}$O$_{3}$ , Co$_{3}$O$_{4}$ at 1473 K. Phase purity was confirmed by X-Ray diffraction (XRD) using Cu K$\alpha$ ($\lambda$ = 1.54056 \text{\AA}) using Philips 1050 diffractometer. Powder neutron diffraction (NPD) measurements were done at PSI, Switzerland with temperature using wavelengths of $\lambda$ = 1.494 \text{\AA} and $\lambda$ = 1.8857 \text{\AA} for room temperature and low temperature respectively. Fullprof suite program used to refine the diffraction patterns. The temperature and magnetic field dependent magnetisation and transport properties measurements were performed using Squid Quantum Design XL-MPMS, MPMS-3 magnetometer and PPMS. 

\section{Results and discussion}

\subsection{Structural Characterization}
\begin{figure}[h!]
	\begin{center}
		\includegraphics[scale=0.6]{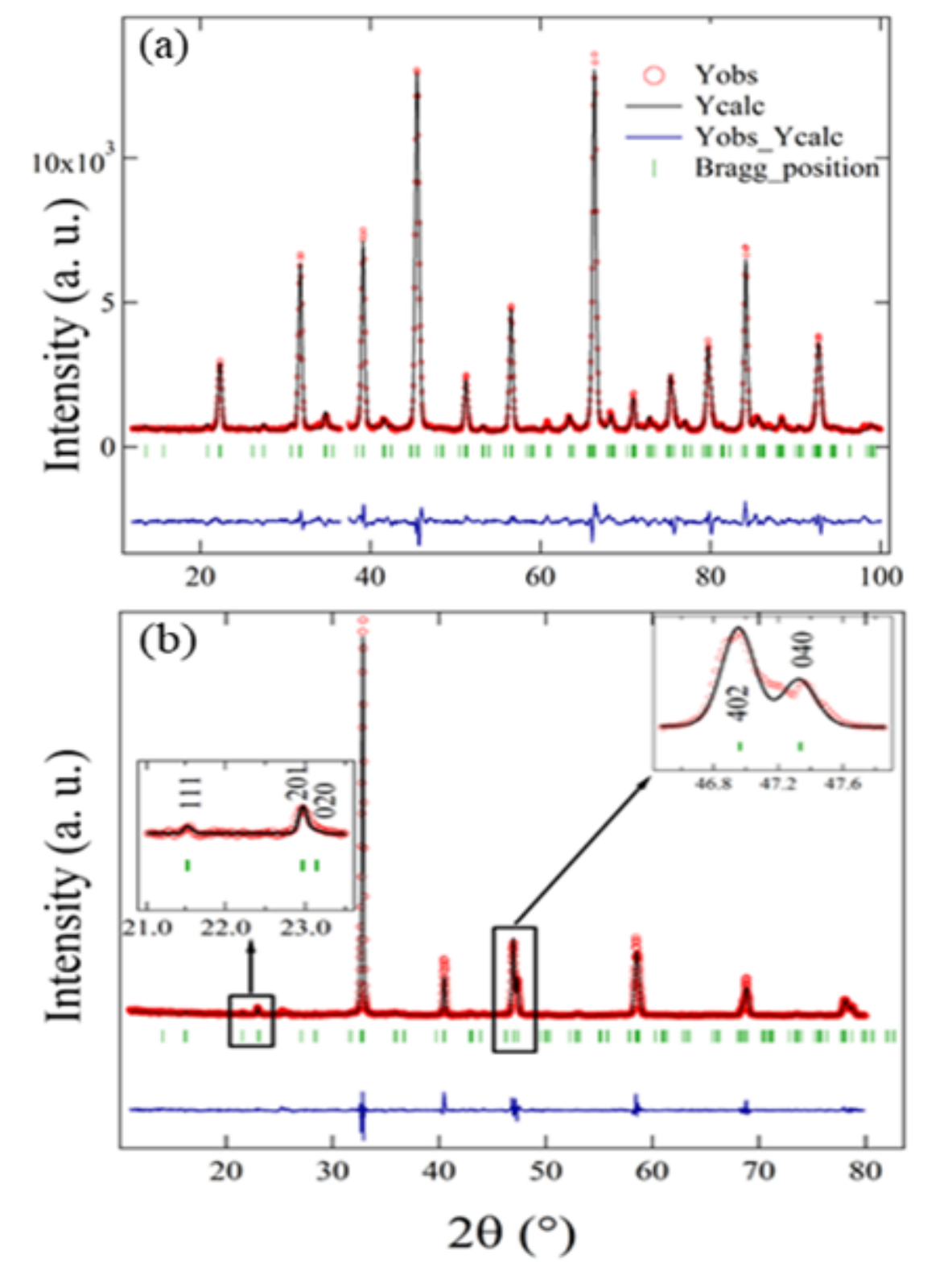}
		\caption{\label{Fig1:1}Final observed, calculated and difference profile (joint Rietveld refinement pattern, Orthorhombic, $\textit{Cmmm}$) of RT XRD and RT NPD patterns for the Sr$_{4}$Fe$_{3}$CoO$_{11}$ sample. Insets in the left figure highlight the peaks characteristics of $\textit{Cmmm}$ space group.}
	\end{center}
\end{figure} 
Preliminary Rietveld refinement of the XRD pattern indicate that the compound can be isolated as single phases, crystallizing in an orthorhombic symmetry, space group (S. G.) $\textit{Cmmm}$ ( $\sim$ 2$\sqrt{2}$a$_{p}$ $\times$ 2a$_{p}$ $\times$ $\sqrt{2}$a$_{p}$). Figure \ref{Fig1:1} shows the final plot of joint Rietveld refinement of the RT XRD and RT NPD pattern of the material. The results obtained from the joint refinement of the RT NPD/RT XRD data, are provided in the supporting information file, and they are similar to the reported one \cite{19}. 

\begin{figure}[h!]
	\includegraphics[width=1.0\columnwidth]{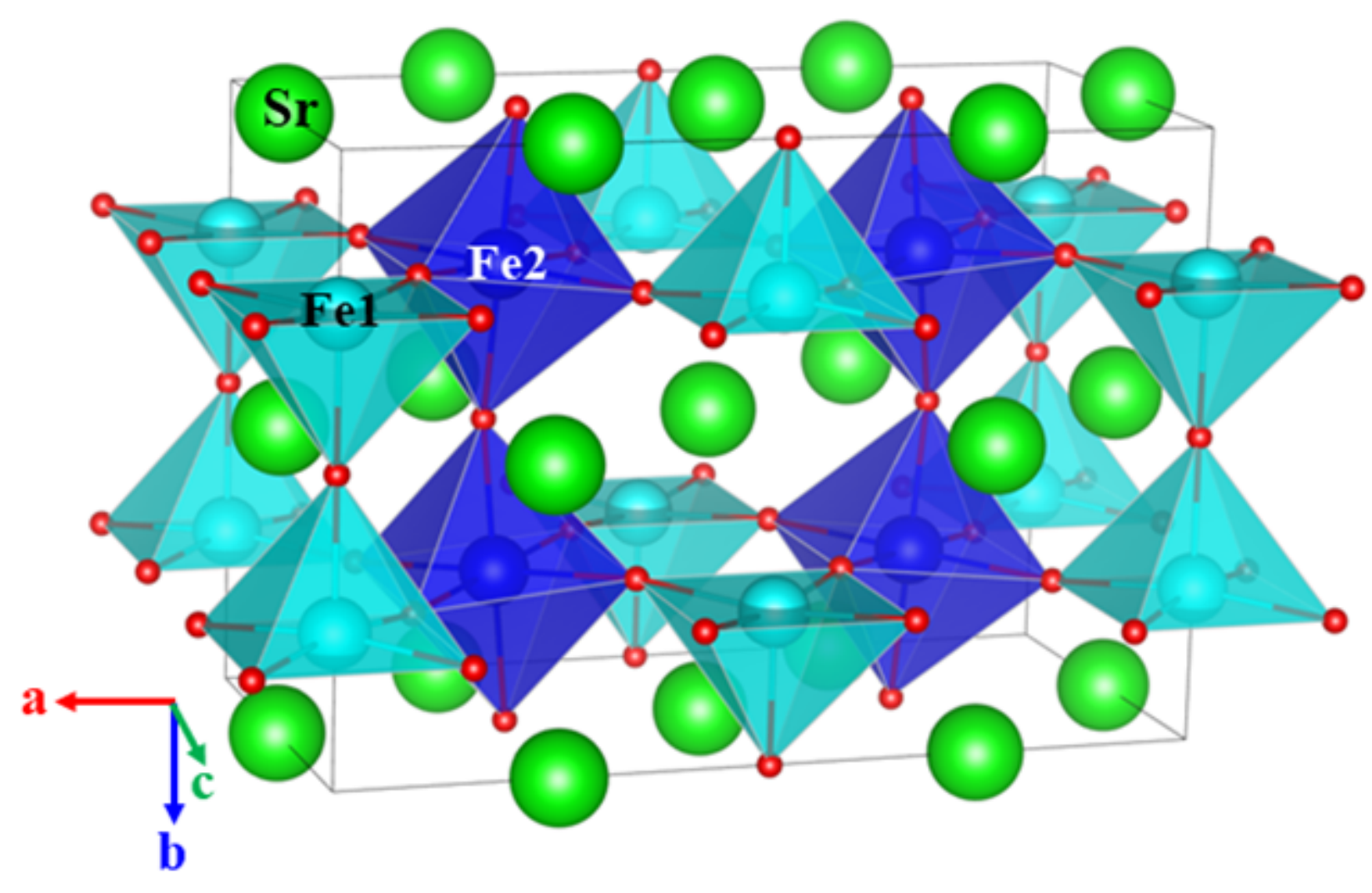}
	\caption{\label{Fig2:2}Crystal structure of  Sr$_{4}$Fe$_{3}$CoO$_{11}$ (Orthorhombic, C$\textit{mmm}$) (oxygen atoms as red spheres).}
\end{figure}

Figure \ref{Fig2:2} shows the crystal strucrure of Sr$_{4}$Fe$_{3}$CoO$_{11}$. Similar to the previous we have found full occupancies for the oxygen positions\cite{19}. Site occupancy refinement indicates a random distribution of Co in both Fe sites. Sr$_{4}$Fe$_{3}$CoO$_{11}$ shares the same vacancy ordered structure of  Sr$_{4}$Fe$_{4}$O$_{11}$.

\subsection{ Magnetic Structure and Magnetic Properties} 
\begin{figure}[h!]
	\begin{center}
		\includegraphics[width=1.0\columnwidth]{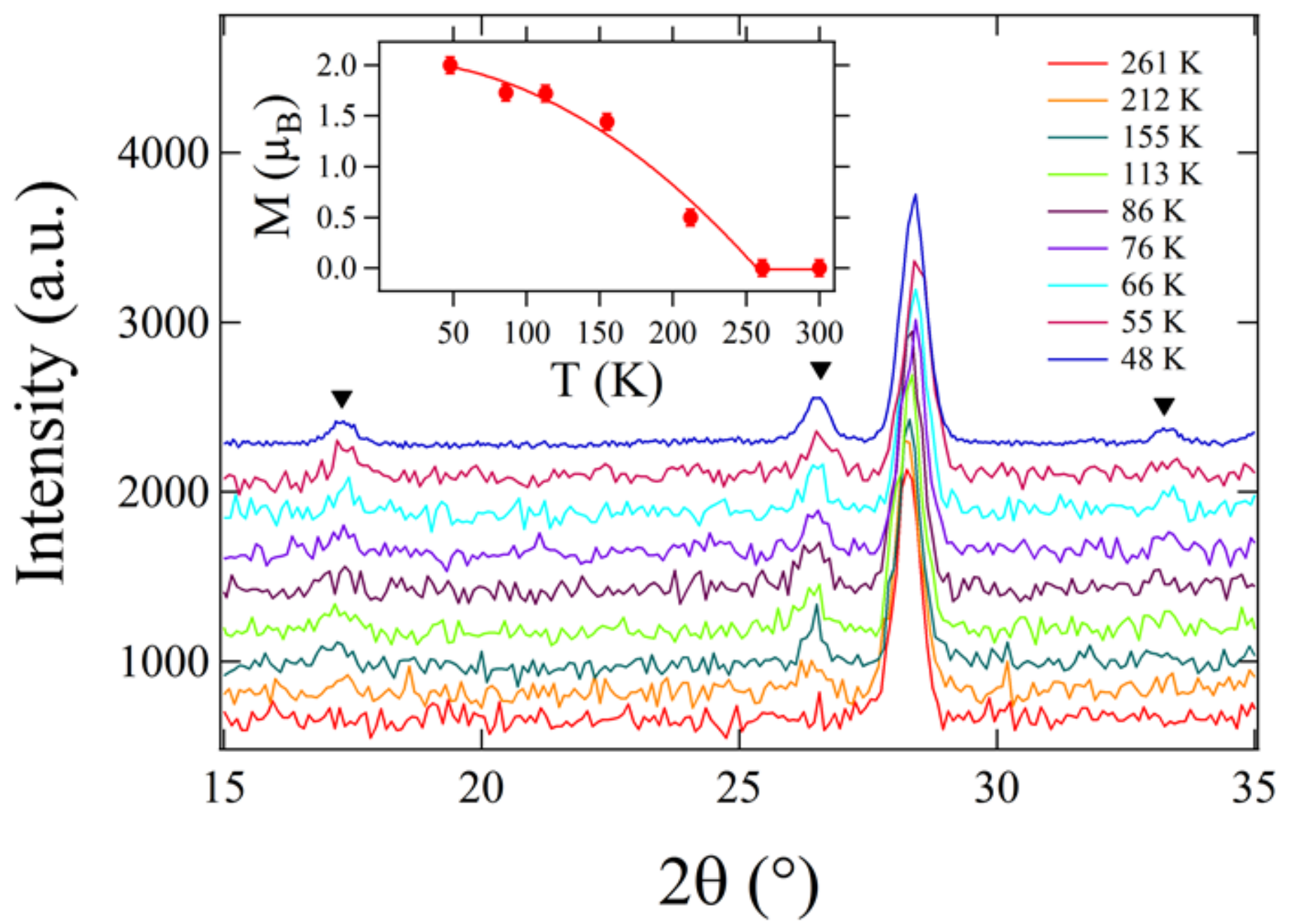}
		\caption{\label{Fig3:3}Neutron powder diffraction patterns collected at different temperatures for Sr$_{4}$Fe$_{3}$CoO$_{11}$ sample. It highlights (by solid triangles) the temperature evolution of the magnetic peaks. Inset shows the temperature dependence of the magnetic moments for the same material.}
	\end{center}
\end{figure}

To get the magnetic structure, NPD patterns were taken down to 48 K. Figure \ref{Fig3:3} shows the temperature evolution of new reflections of magnetic origin for Sr$_{4}$Fe$_{3}$CoO$_{11}$. We have used the Shubnikov space group C$\textit{m'm'm}$ (propagation vector k = 0, 0, 0, $\Gamma$$_7$) to refine the low temperature NPD patterns of the present material. It shows a G-type AFM ordering below 250 K. The temperature evolution of the magnetic moments is shown as an inset in figure \ref{Fig3:3}.

\begin{figure}[h!]
\begin{center}
	\includegraphics[width=1.0\columnwidth]{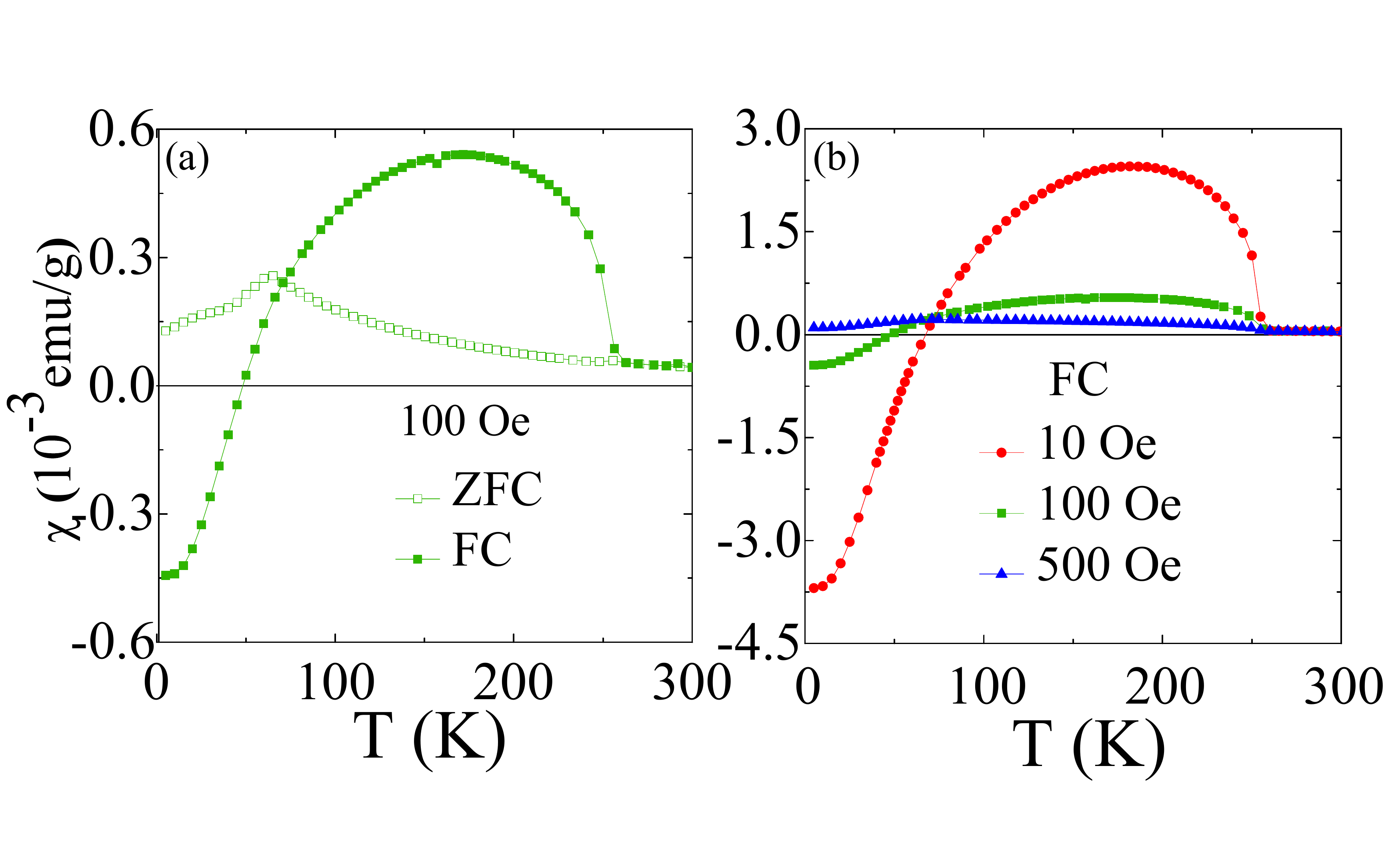}
		\caption{\label{Fig4:4}Temperature dependence (a) FC and ZFC magnetic susceptibility curves measured at 100 Oe and (b) FC susceptibility curves measured under different magnetic field for Sr$_{4}$Fe$_{3}$CoO$_{11}$. It highlights the dependence of magnetization reversal (negative magnetization) with magnetic field.}
	\end{center}
\end{figure}

Figure \ref{Fig4:4}(a) shows the temperature dependence of the FC and ZFC susceptibility ($\chi$) collected at 100 Oe for Sr$_{4}$Fe$_{3}$CoO$_{11}$. An AFM ordering (T$_{N}$ = 255 K $\pm$ 2 K) and then a large irreversibility between the FC and ZFC susceptibility is apparent in the figure. The sudden rise of the FC susceptibility curve below T$_{N}$ highlights the existence of weak ferromagnetic (WFM) correlations below the AFM transition temperature. Below T$_{N}$, the FC susceptibility curve bifurcate from the ZFC with a broad maximum at $\sim$ 165 K $\pm$ 3 K and then on further cooling, it continuously dropping towards a negative value by crossing the $\chi_{ZFC}$ at T$_{Cross}$ $\sim$ 70 K $\pm$ 4 K and reaches the magnetic compensation point ($\chi_{FC}$ = 0) at T$_{Comp}$ = 47 K. 

Below this compensation temperature, the FC susceptibility is negative down to the lowest measured temperature (2 K), i.e., the resultant magnetization is opposite to the applied magnetic field. Thus, the FC magnetization exhibit temperature induced magnetization reversal phenomenon. The ZFC susceptibility curve shows a peak at $\sim$60 K, indicates a complex magnetic interaction/magnetic frustration at that temperature. To see the characteristics of negative magnetization observed in the sample, we have measured the magnetization at different applied magnetic fields. Figure \ref{Fig4:4}(b) shows the temperature dependence of FC susceptibility data collected under different magnetic field. It shows that the compensation temperature is shifted towards the lower temperature side with increasing magnetic field and when the applied magnetic field increases to 500 Oe, the FC susceptibility switches to the positive side. This suggests that the net magnetization, originally opposite to the direction of the applied magnetic field is oriented along the direction of the applied magnetic field at a higher magnetic field. It is important to recall here that the isostructural compounds, such as Sr$_{4}$Fe$_{4}$O$_{11}$ and the samples with lower Co concentration do not show any magnetization reversal \cite{12,13,18}. In this structure, there are two different Fe sites, square pyramidal Fe1 sites and octahedral Fe2 sites (see figure \ref{Fig2:2}). The Fe2 sub lattice form a long range ordered AFM spin structure \cite{11,17}. 
Each Fe1 will have two Fe2 neighbors with up-spin and two with down spin . As a result, the Fe1 magnetic moments are topologically frustrated due to a ferromagnetic Fe1-O-Fe2 double exchange and AFM Fe1-O-Fe1 super-exchange interactions. Also, in the present material Co and Fe ions are randomly sharing two different crystallographic (Fe) sites with a net AFM interaction. Therefore, the doped cobalt cations also give rise to additional canting of Fe moments; this in fact is observed by the sudden rise of $\chi_{FC}$ below T$_{N}$. In low symmetric antiferromagnetic materials, the existence of WFM is mainly due to the single-ion magnetocrystalline anisotropy or DM interactions or both \cite{20, 21}. Also, in certain cases, where the net moment caused by these mechanisms can be oriented in the opposite field direction \cite{20,21,22}, and they can have different temperature dependencies. Under this circumstances, negative magnetization can arise under low applied magnetic fields. In such a case, the observed magnetization reversal in the present sample can be attributed to the increase of the magnetocrystalline anisotropy with Co doping. As we reach the 25$\%$ substitution level (Sr$_{4}$Fe$_{3}$CoO$_{11}$), the magnetocrystalline anisotropy dominates and the net magnetic moment is aligned opposite to the applied magnetic field.

\subsection{Exchange Bias and Training effect} 

\begin{figure}[h!]
	\begin{center}
		\includegraphics[width=1.0\columnwidth]{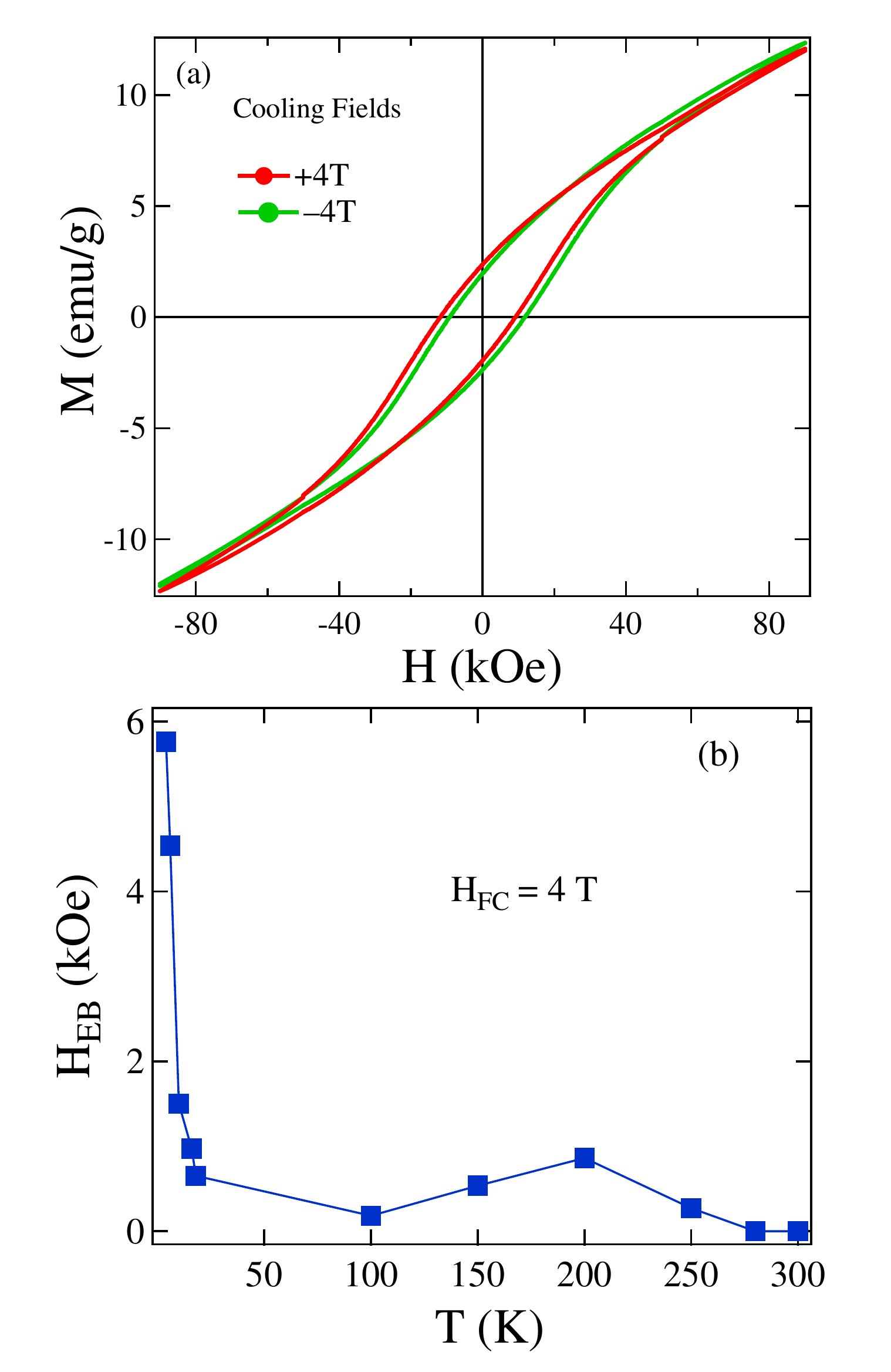}
		\caption{\label{Fig5:5}(a) Magnetic field dependent magnetization loop (M-H) at 10 K measured in field
			cooling mode and (b) temperature dependence of the exchange bias field (H$_{EB}$) for Sr$_{4}$Fe$_{3}$Co
			O$_{11}$. A large value of exchange bias field $\sim$ 1.5 kOe is observed at 10 K.}
	\end{center}
\end{figure}

Recently, the complex magnetic materials have attracted remarkable interest for exploring the exchange bias effect (EB) \cite{4,23,24,25,26}. The EB phenomenon is generally demonstrated as a shift \cite{27} in the isothermal magnetization loop with respect to the field axes. It can be utilized in several technological applications such as magnetic recording read heads \cite{28}, random access memories \cite{29} and other spintronic devices \cite{30,31}. Figure \ref{Fig5:5}(a) shows the magnetic hysteresis loop (M-H) between $\pm$90 kOe for the Sr$_{4}$Fe$_{3}$CoO$_{11}$) at 10 K, measured in FC condition (cooling field, H$_{FC}$ = $\pm$40 kOe). For positive cooling field (H$_{FC}$ = 40 kOe), the M-H loop is shifted towards the left field axis and it is opposite to the M-H loop measured in negative cooling field (H$_{FC}$ = -40 kOe) at 10 K. This clearly indicates that the observed EB effect has not emerged from the unsaturated minor loop. The EB field is calculated from the M-H loop shift ((H$_{EB}$ = -(H$_{C(L)}$ + H$_{C(R)}$)/2, where H$_{C(L)}$ and H$_{C(R)}$ are the left and right intercepts with the field axis). A very high value of H$_{EB}$ $\sim$ 1.5 kOe is observed at 10 K. The suitable cooling field was chosen by measuring the M-H loops in various cooling fields (H$_{FC}$) ranging from 0.5 kOe to 60 kOe at 4 K. To study the nature of exchange bias properties, and to know its origin, we have measured the temperature variation of the exchange bias effect. Figure 5(b) shows the temperature dependence of the exchange bias field (H$_{EB}$). The EB effect is started to emerge from the AFM ordering temperature (H$_{EB}$ = 271 Oe at 250 K). Then, at 25 K the H$_{EB}$ shows a sharp rise and shows a very high value of 5.8 kOe at 4 K.

Sr$_{4}$Fe$_{3}$CoO$_{11}$ shows training effect. In training effect, the uncompensated spin at the interface shows relaxation due to the repetition of the M-H measurement. As a result the exchange bias field shows variation on the number of successive loops studied. Figure 6 highlights the variation of H$_{EB}$ with number of field cycles (n). In this instance, the exchange bias field shows a decreasing tendency as the cycle number (n) increases. The decrease in H$_{EB}$ can be
described by fitting the data to the following empirical power law for n $\geq$ 2 \cite{32}.
\begin{equation}
H_{EB}(n)-H_{E\infty} = k/\sqrt{n}  ,
\label{eqn1:1}
\end{equation}

Where H$_{EB}$ (n) = exchange field in the nth cycle. H$_{E\infty}$ = exchange field for n = ${\infty}$, and k is a system dependent constant. The obtained value of H$_{E\infty}$ is 0.3 kOe, the remnant H$_{EB}$ at 10 K. Also, our experimental data follow the recursive formula suggested by Binek \cite{33} i.e.,  
\begin{equation}
H_{EB}(n+1)-H_{EB} (n) = \gamma (H_{EB}(n)-H_{E\infty})^{3}  ,
\label{eqn2:2}
\end{equation}                   																								                                                   
where $\gamma$ = 1/2k$^{2}$. Notably, the Eq. 1 is applicable for data starting from n = 2, while Eq. 2 is also valid for for n = 1.                                              
\begin{figure}[h!]
	\begin{center}
		\includegraphics[width=1.0\columnwidth]{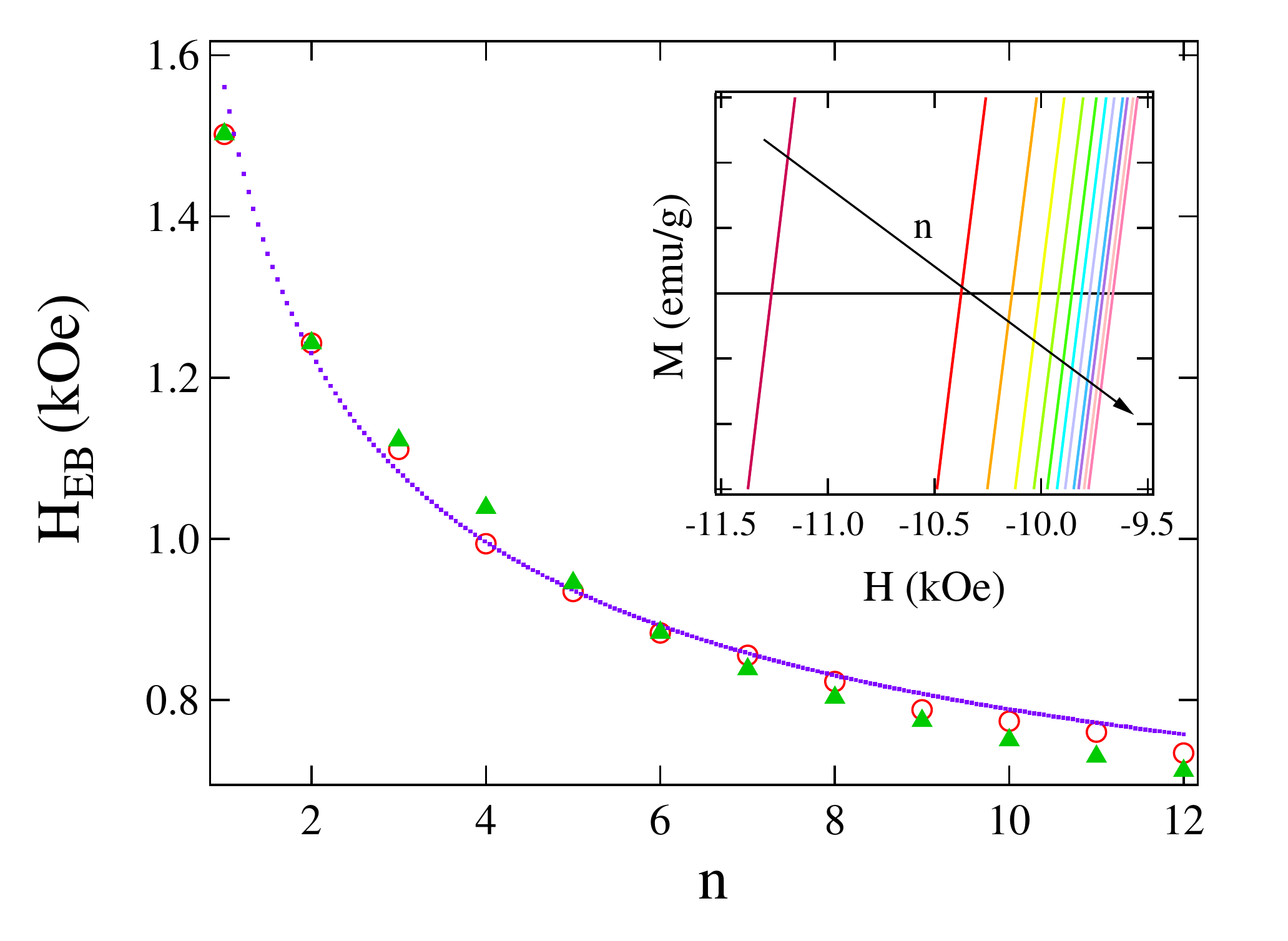}
		\caption{\label{Fig6:6}The exchange bias field (H$_{EB}$) vs number of field cycle (n) (red open circles) as calculated from the training effect magnetic hysteresis loops (M-H) at 10 K. Fitting of Eqn.
			\ref{eqn1:1} (see the text) to the H$_{EB}$ data is shown as dotted line. The data points generated using the recursive formula, Eqn. \ref{eqn2:2} (see the text) is shown by green triangles. An enlarged part of the consecutive M-H loops (left intercept with the field axis) is shown in the inset.}
	\end{center}
\end{figure}
As discussed previously, the Fe1 site is topologically frustrated due to a ferromagnetic Fe1-O-Fe2 double exchange and AFM Fe1-O-Fe1 super-exchange interactions. Also, the random distribution of the Co cation in two different Fe sites in having a net AFM interaction originates additional frustration in the sample. Therefore, the giant EB effect in this material could be due to the existence of different magnetic interactions in the layered structure. 
\subsection{Magnetoresistance} 
\begin{figure}[h!]
	\begin{center}
		\includegraphics[width=1.0\columnwidth]{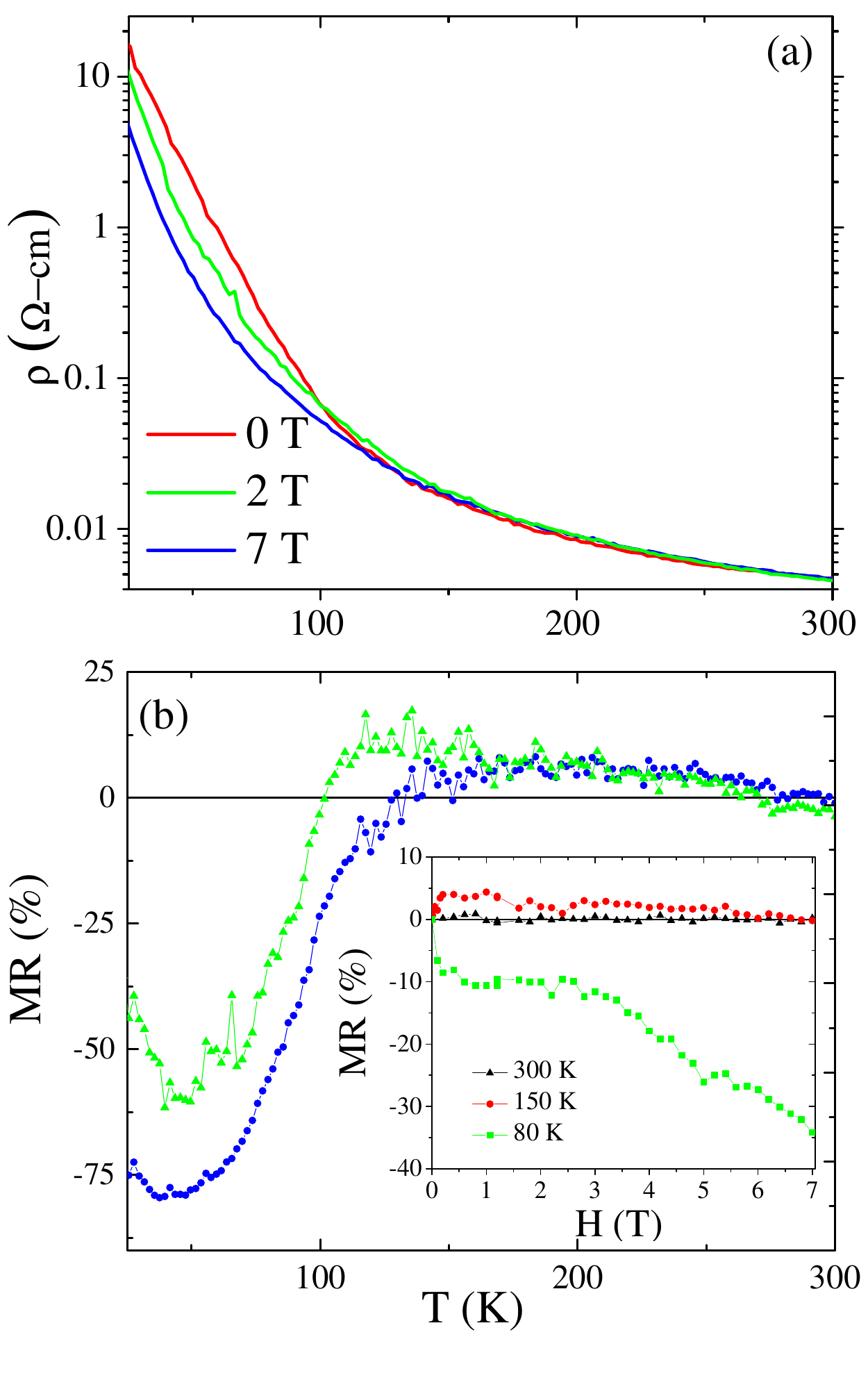}
		\caption{\label{Fig7:7}(a) Temperature dependence of resistivity measured at different magnetic field. (b) Temperature dependence of magnetoresistance at different magnetic field for Sr$_{4}$Fe$_{3}$CoO$_{11}$. Magnetic field dependence of magnetoresistance at different fixed temperature is shown in the inset in figure (b).}
	\end{center}
\end{figure}

Figure \ref{Fig7:7}(a) shows temperature dependence of the electrical resistivity [$\rho(T)$] at different magnetic fields. All curves exhibit a semiconducting-type behavior. We do not observe any change in the resistivity data measured in the presence of a magnetic field down to 250 K. However, below 250 K $\rho(T)_{H}$ start to deviate from $\rho(T)_{H=0kOe}$ and becomes higher down to 100 K. Then, with further lowering the temperature, an interesting crossover from positive to negative magnetoresistance is observed. The temperature variation of magnetoresistance (MR) (MR = [$\rho(T)_{H}$-$\rho(T)_{H=0kOe}$]/$\rho(T)_{H=0kOe}$) is highlighted in figure \ref{Fig7:7}(b). We have observed a huge value of negative MR over a wide temperature range (100 - 20 K), and at 25 K the MR is -80$\%$. Above 100 K, the magnetoresistance turns into the positive side ($\sim$ 10$\%$ at 150 K). In order to confirm the MR behavior and its crossover, resistivity ($\rho$(H)) was measured in the presence of magnetic field at different fixed temperatures. Similar to the $\rho(T)$ measurements at different magnetic fields, it confirms the crossover of the magnetoresistance. However, the small positive magnetoresistance observed above 100 K is found to decrease with increasing magnetic field. The opening up of the band gap due to the dominant antiferromagnetic interactions at higher temperature could be the origin of the positive magnetoresistance in this sample. A similar situation is already observed in a few iron based compounds \cite{34,35}. Then with further lowering the temperature, magnetic frustration increases and this develops the spin-dependent scattering of carriers which in turn originate the conventional (negative) MR in this sample. At a lower temperature, the magnetic disorder due to the existence of different magnetic interactions in the layered structure enhances the spin-dependent scattering of the carriers and therefore the electrical resistance in zero magnetic field.

\section{Conclusions}
We have investigated the structural, magnetic, exchange bias and magneto-transport properties for the oxygen deficient Sr$_{4}$Fe$_{3}$CoO$_{11}$ materials. The compound shows orthorhombic symmetry, crystallizing in the C${\textit{mmm}}$ space group, isostructural with Sr$_{4}$Fe$_{4}$O$_{11}$. The onset of the antiferromagnetic (G-type) transition is observed at T$_{N}$ = 255 K. Then, with further lowering the temperature, the material shows a range of exciting physical properties. Temperature induced magnetization reversal is observed at a compensation temperature (T$_{Comp.}$) of 47 K. The magnetization reversal is observed due to the increased magnetocrystalline anisotropy with Co substitution. Also, two technologically important phenomena, magnetoresistance and exchange bias is observed near room temperature in this compound. The exchange bias is observed below the AFM ordering temperature when the sample is cooled with a magnetic field. The large exchange bias effect in this sample is attributed to the magnetic frustration in the structure. The magnetoresistance starts to appear below T$_{N}$ = 255 K and reaches a huge value of 80$\%$ at 25 K in 7 T. An interesting crossover in the magnetoresistance value from negative to positive side is observed $\sim$ 100 K. The opening up of the band gap due to the dominant antiferromagnetic interactions at higher temperature could be the origin of the positive magnetoresistance in the sample. On the other hand, increasing magnetic frustration in the structure with lowering the temperature develops the spin-dependent scattering of carriers which originate the conventional (negative) magnetoresistance.

\section*{Acknowledgements}
PM and SM have contributed equally. SM, CM and OT thank CNRS, Universit\'e Bordeaux, and SOPRANO project (Seventh Framework Programme FP7/ 2007-2013 under Grant Agreement No. 214040) for funding this work. R. P. S. acknowledges Science and Engineering Research Board (SERB), Government of India for the Ramanujan Fellowship through Grant No. SR/S2/RJN-83/2012. S. M acknowledges Science and Engineering Research Board, Government of India for the NPDF fellowship (Ref. No. PDF/2016/000348). This work is based on experiments performed at the Swiss spallation neutron source SINQ, Paul Scherrer Institute, Villigen, Switzerland. The authors also thank Dr. Vladimir Pomjakushin for his help in collecting the neutron powder diffraction patterns.

\end{document}